\documentclass[emulateapj]{emulateapj}
\usepackage{times,pslatex}
\usepackage{amsmath,amsfonts,graphicx}
\usepackage{auto-pst-pdf}
\usepackage{epsfig}
\usepackage{rotate}
\usepackage{color}

\newcommand{\hst}{{\it HST}}
\newcommand{\um}{\mbox{$\mu$m}}
\newcommand{\bff}{}

\slugcomment{Accepted for publication in the Astrophysical Journal}
\shorttitle{Transit of HD 189733b}
\shortauthors{McCullough et al.}

\received{2014 Feb 27}
\accepted{2014 Jun 20}
\begin{document}

\title{Water Vapor in the Spectrum of the Extrasolar Planet HD 189733\lowercase{b}: 1. the Transit}

\author{P.~R.~McCullough\altaffilmark{1,2}}
\author{N.~Crouzet\altaffilmark{1,3}}
\author{D.~Deming\altaffilmark{4,5}}
\author{N.~Madhusudhan\altaffilmark{6,7}}

\affil{\altaffilmark{1} Space Telescope Science Institute, Baltimore, MD 21218, USA}
\affil{\altaffilmark{2} Department of Physics and Astronomy, Johns Hopkins University, 3400 North Charles Street, Baltimore, MD 21218}
\affil{\altaffilmark{3} Dunlap Institute for Astronomy \& Astrophysics, University of Toronto, 50 St. George Street, Toronto, Ontario, Canada M5S 3H4}
\affil{\altaffilmark{4} Department of Astronomy, University of Maryland, College Park, MD 20742, USA}
\affil{\altaffilmark{5} NASA Astrobiology Institute's Virtual Planetary Laboratory}
\affil{\altaffilmark{6} Yale Center for Astronomy \& Astrophysics, Yale University, New Haven, CT 06511, USA}
\affil{\altaffilmark{7} Institute of Astronomy, University of Cambridge, Madingley Road, Cambridge, CB3 0HA, UK.}

\email{pmcc@stsci.edu}

\begin{abstract}
We report near-infrared spectroscopy of the gas giant planet HD
189733b in transit.  We used the {\it Hubble Space Telescope} Wide
Field Camera 3 (\hst\ WFC3) with its G141 grism covering 1.1 \um\
to 1.7 \um\ and spatially scanned the image across the detector
at 2\arcsec$s^{-1}$. When smoothed to 75 nm bins, the local maxima
of the transit depths in the 1.15 \um\ and 1.4 \um\ water vapor features
respectively are $83\pm 53$ ppm and $200\pm 47$ ppm greater than
the local minimum at 1.3 \um. We compare the WFC3 spectrum with the composite
transit spectrum of HD 189733b assembled by \citet{PON13}, extending
from 0.3 \um\ to 24 \um. Although the water vapor features in the
WFC3 spectrum are compatible with the model of non-absorbing,
Rayleigh-scattering dust in the planetary atmosphere \citep{PON13}, we also
re-interpret the available data with a clear planetary atmosphere.
In the latter interpretation, the slope of increasing transit depth
with shorter wavelengths from the near infrared, through the visible
and into the ultraviolet is caused by unocculted star spots, with
a smaller contribution of Rayleigh scattering by molecular hydrogen
in the planet's atmosphere. At relevant pressures along
the terminator, our model planetary atmosphere's temperature is $\sim$700
K, which is below the condensation temperatures of sodium- and
potassium-bearing molecules, causing the {\bff broad wings of the spectral lines} of Na I and K I at 0.589 \um\ and 0.769
\um\ to be weak.
\end{abstract}


\section{Introduction}
\label{sec:intro}

{\bff
The K1V star HD 189733 hosts the transiting gas giant planet HD 189733b, at an orbital
distance of 0.03 AU \citep{BOU05}. In some ways this system 
is nearly optimally suited for study with transmission spectroscopy. Its short planetary
orbital period (2.2 days), its great stellar brightness
(K = 5.5 mag), and its large planet-to-star radius ratio (0.15) combine for
convenient observations of high sensitivity and high contrast.
On the other hand, a significant negative is its intrinsic photometric variability,
$\sim2$\% peak to valley, due to rotational modulation of star spots \citep{AIG12}.
As such, it is photometrically more variable than $\sim$95\% of stars in the Kepler
sample (\citealp{MCQ14}; \citealp{CIA11}; cf. also Figure 5 of \citealp{MCQ12}). As this paper will
discuss, star spots complicate the interpretation of transmission spectra, including
those of HD 189733b.
}

Transmission spectra provide a powerful means to detect molecular
species in exoplanetary atmospheres due to the long path length of
the stellar light through the planetary atmosphere (e.g. \citealp{BRO01};
\citealp{SEA00}; \citealp{FOR10}). The differential transit depth of
a molecular feature is
directly proportional to the scale height of the atmosphere.
For hot Jupiters, with high temperatures and low mean
molecular weight (i.e. H$_2$-rich), the differential transit depth in specific molecular
bands is large enough to be observable with existing instruments.
{\bff In particular, water vapor has bands with large cross sections observable in the near infrared.}

Upper limits or marginal detections have been reported of water in transmission for exoplanets 
HD 209458b \citep{BAR07}, XO-1b (\citealp{TIN10}; \citealp{CRO12}), and XO-2b \citep{CRO12}.
From {\it Spitzer} IRAC transit photometry at 3.6 $\mu$m,
5.8 $\mu$m, and 8 $\mu$m, \citet{TIN07} reported evidence of water vapor in the atmosphere of HD 189733b (but cf.
\citealp{DES09}).
The near-infrared transmission spectrum of HD 189733b in particular has garnered considerable attention, with the same NICMOS data
being re-analyzed multiple times, with \citet{SWA14} being the latest. {\bff The latter meta-analysis
of the transmission spectra of \citet{SWA08}, \citet{GIB11}, and \citet{WAL13}, predicted the amplitude
of HD 189733b's 1.4 \um\ water vapor feature to be 300 ppm to 400 ppm.} Other meta-analyses have been
performed also (e.g. \citealp{MAD09}). Similarly, from ground-based {\it IRTF} 
near-infrared transmission spectra of HD 189733b, \citet{DAN14} reported water vapor features including ones
at 1.15 \um\ and 1.4 \um\ with amplitudes at least twice those reported here.

Significant improvement has come with the \hst\ instrument WFC3.
Recently, \citet{MAN13} reported WFC3 observations of transits of WASP-12b, WASP-17b, and WASP-19b. \citet{HUI13}
reported WFC3 observations of WASP-19b alone.
Made under \hst\ program 12181 (P.I. Deming), those observations preceded the development of spatial scanning
and like prior observations, produced marginal detections or upper limits on the 1.4 $\mu$m water vapor feature.
Similar WFC3 spectra have been reported for WASP-12b observed in staring mode under \hst\ programs 12230 (P.I. Swain; \citealp{SWA13})
and 12473 (P.I. Sing; \citealp{SIN13}).
Coupling WFC3 to the new technique of spatially scanning the Hubble Space Telescope (\hst) has
enabled detections of water vapor in transmission spectra for HD 209458b, XO-1b, and HAT-P-1b
under \hst\ program 12181 (P. I. Deming) and now HD 189733b (program 12881; P. I. McCullough).

{\bff For some planets, clouds and/or haze can greatly affect our view of water
vapor even if it exists in their atmospheres. Such is the case apparently
for the smaller exoplanets GJ 1214b \citep{KRE14a} and GJ 436b
\citep{KNU14} that both exhibit featureless, flat spectra. Planets with no atmospheres, or insufficiently
large scale heights of their atmospheres, would also present undetectably weak
diagnostic spectral features in their atmospheres \citep{DEM10}.
Even if molecular features are observed in a clear atmosphere, a degeneracy
exists between the composition and the characteristic temperature of the region
probed by the data, since the scale height is proportional to
temperature. High resolution (R${\bff >50}$), high sensitivity data can
place joint constraints on the composition and temperature simultaneously
(e.g. \citealp{BEN12}).
}

Due to their geometry of light rays grazing the terminator region \citep{FOR05a},
transmission spectra are more sensitive to clouds/haze
and/or small atmospheric scale heights than day-side emission spectra, which have
their own limitations. \citet{GRI08} attributed to water a 10 \um\ absorption feature in the
emission spectrum of HD 189733b, and recently, \citet{BIR13} have done so 
at 3.2 \um\ using phase curves rather than observations at secondary eclipse. Such measurements
indicate potential for comparison of spectra obtained by the complimentary
techniques of transits, eclipses, and phase curves (cf. \citealp{KRE14b} and \citealp{STE14}).

The spectral observations reported here were obtained in spatial
scanning mode \citep{MCC12a} with the \hst\ in order to rapidly
collect a large number of photons from HD 189733 without it also
saturating the detector. In scanning mode, the \hst\ scans
across a small segment of a great circle on the sky while the
detector is integrating photo-generated charge from objects as they trail
across the detector.  Since the first such observation of a transiting
exoplanet, GJ 1214b on April 18, 2011 (program 12325, P.I. MacKenty; \citealp{MCC12b}),
spatial scanning has become the nominal method of spectroscopic
observation of bright exoplanet host stars.  \citet{DEM13}
scanned one transit each of HD 209458b and XO-1b; \citet{KRE14a}
and \citet{KNU14} each scanned multiple transits of GJ 1214b
and GJ 436b respectively. \citet{WAK13} scanned one transit of HAT-P-1b.
\hst\ program 12181 (P.I. Deming) recorded one transit of HD 189733b using spatial scans.
However, the Earth occulted the telescope's
view of the star during the main part of the transit event between contacts 2 and 3. 

\citet{GIB12} observed transits of HD 189733b in the near-infrared with WFC3,
under program 11740 (P.I. Pont), although they did so prior to the advent of spatial scanning on
\hst. In those observations, the very bright star saturated the detector even
before each exposure began, because the detector does not have a mechanical
shutter.
They salvaged light curves from the insensitive blue and red ends of the spectra.
Based upon comparison of those measurements at $\sim$1.1 \um\ and $\sim$1.6 \um\ 
with measures by \citet{PON08} between 0.5 \um\ and 1.0 \um, Gibson et al. tentatively
suggested that haze dominates the transmission spectrum of HD 189733b
both in the visible and the near-infrared. 
However, they concluded that future observations of the sort reported here would provide the
evidence to resolve the issue of whether exoplanetary hazes or clouds obscure
spectral features in the near-infrared.

This paper is organized as follows. In Section \ref{sec:obs} we
describe the observations, which we analyze in Section \ref{sec:analysis}.
Results are in Section \ref{sec:results}. We separately discuss star spots
(\S\ref{sec:spots}) and the exoplanet's atmospheric spectrum (\S\ref{sec:planet}), and
the combination of the two (\S\ref{sec:combined}).
We draw conclusions and summarize in Section \ref{sec:summary}.

\section{Observations}
\label{sec:obs}

\hst\ program 12881 (McCullough, P.I.) was allocated five HST orbits
to observe the transit of HD 189733b and five orbits to observe its eclipse. The
eclipse observations are reported by \citet{CRO14}. In both
cases, the event (transit or eclipse) occurs in the fourth of the
five HST orbits. By design and common practice, the data from the
first HST orbit is considered unreliable, as the telescope settles
to its new thermal environment associated with a new orientation
to the Sun after slewing to the target, and charge traps in the HgCdTe
detector equilibrate to the unusual voltage associated with strong illumination.
Discarding the data from
the first orbit would leave two reliable HST orbits pre-transit, one
in-transit, and one post-transit.\footnote{Of the two separate analyses (\S\ref{sec:analysis}), the N. C. one
discarded data from the entire first orbit whereas the D. D. one discarded only the first eight scans
of the first orbit.}

We observed HD 189733b in transit with HST WFC3 on June 5, 2013 (Table \ref{tbl-1}).
We used the G141 grism, to obtain slitless spectroscopy with spectral
coverage from 1.1 \um\ to 1.7 \um\ at a resolution R = $\lambda /
\Delta\lambda = $130 and a dispersion of 4.7 nm pixel$^{-1}$ \citep{DRE14}.
In each HST orbit of the visit, we obtained a set of thirty-two 5.97-s exposures,
each with the RAPID sample sequence of seven samples, and each with
a 512 pixel by 512 pixel subarray.  As noted in the Introduction,
the spectra were obtained in spatial scanning mode. We interleaved
scanning first forward and then reversing direction, obtaining
sixteen forward-reverse pairs of exposures in each orbit.  We had
intended to locate the target on the detector such that its first-order
spectrum would be entirely within a region read from a single
amplifier; however we miscalculated and the 1st order spectrum
crosses an amplifier boundary (at physical column 512) corresponding
to $\lambda = 1.55$ \um.  A similar miscalculation caused the spectrum
to scan off the top of the subarray and hence some of the samples
were lost.\footnote{We adjusted the position for
the eclipse observations \citep{CRO14}.}

\begin{figure}
\plottwo{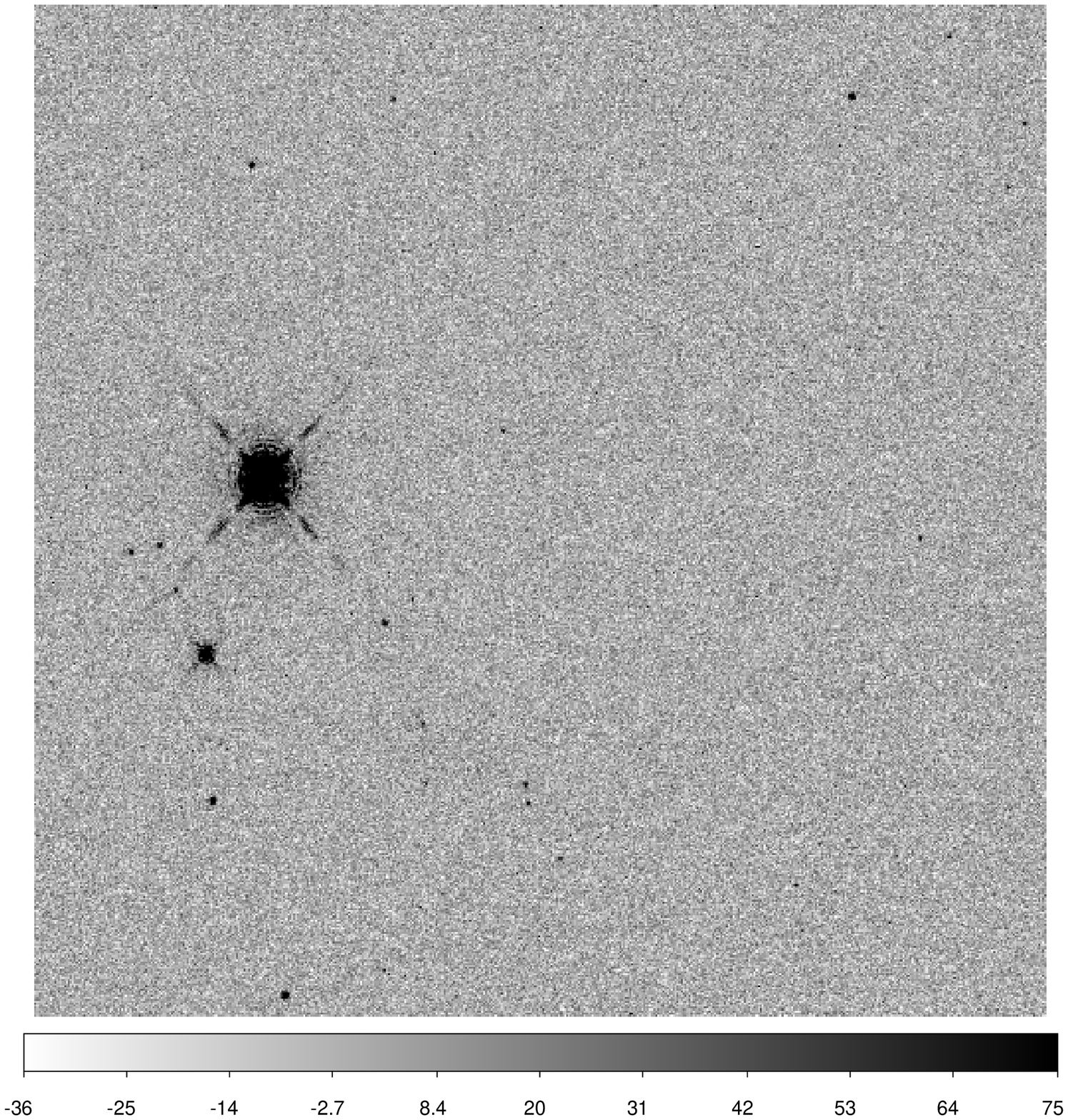}{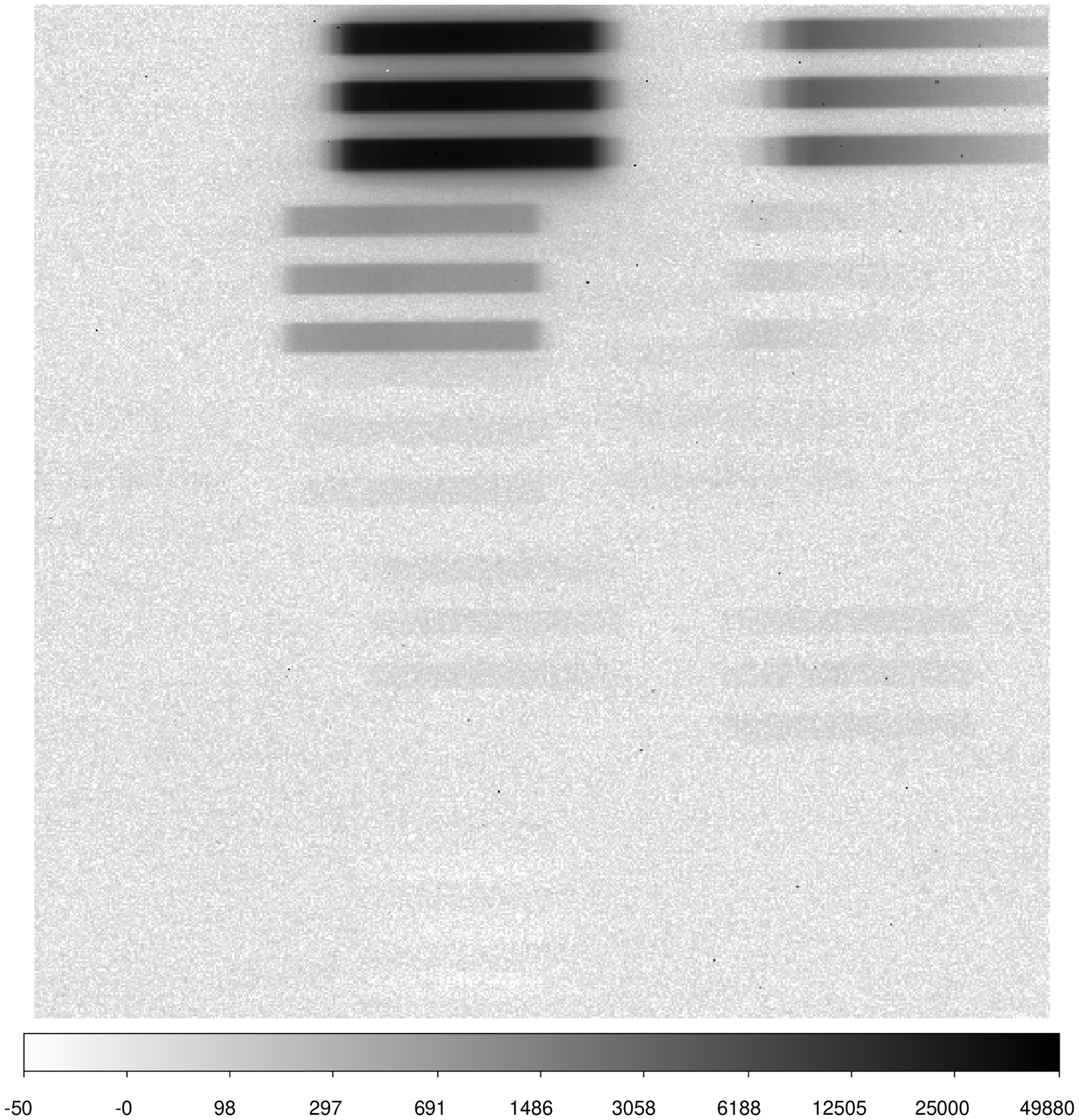}
\caption{(Left) Direct image of HD 189733, from a 2.6-second exposure taken through filter F132N at the start of the visit. 
North is approximately to the right and East approximately up. The secondary star HD 189733B is at 7 o'clock with respect to
the primary HD 189733A. We oriented HST so as to separate the two stars vertically on the detector in order to keep
their two scanned spectra separate in our analysis.
(Right) Scanned spectra of HD 189733. 
The orientation and pixel scale of this image are the same as the direct image.
A logarithmic grey scale is indicated at the bottom in units of e$^-$ pixel$^{-1}$. 
For illustration purposes only, we numerically superposed three sub-exposures corresponding to the 3rd, 5th, and 7th differences between adjacent RAPID
samples; the 4th and 6th differences (not shown) fill the gaps. 
Three first-order spectra of the primary star HD 189733A appear as black rectangles at middle top. 
Three first-order spectra of the secondary star HD 189733B appear in grey below the triad of primary star spectra.
To the right of each first order spectrum, a corresponding 2nd order spectrum appears; they were not analyzed.
With the 512 pixel by 512 pixel subarray and the selected sample sequence, each difference between samples corresponds to
0.853 s exposure and to 1.7 arcsec (14 pixels) vertically when scanned at 2 arcsec s$^{-1}$.
\label{fig1}}
\end{figure}

In the forward direction, samples 1 through 5 occur when HD 189733's
spectrum is on the subarray but during samples 6, 7, and 8 the
spectrum is off the subarray and hence only the first four differences
between samples are useful.  In the reverse direction, samples 3
through 8 recorded HD 189733's spectrum, and samples 1, and 2 were
lost (Figure \ref{fig1}). The loss of 5 differences between samples of 14 total implies
a loss in potential, Poisson-limited, signal-to-noise ratio of 20\%.

We also obtained a direct image as the first exposure of the visit
(Figure \ref{fig1}); the purpose of the direct image is to provide
a high-resolution image to verify the location of the target with
respect to any potentially contaminating stars in the same field
of view.  HD 189733 is a double star; the secondary is 4.0 magnitudes
fainter in H band and is 11.2 \arcsec\ away from the primary at a
PA of 246 \arcdeg\ \citep{BAK06}.  There are a few stars
closer to HD 189733A in projection than HD 189733B, but they are
much fainter; in H band each is 1\% to 2\% of the brightness of HD
189733B, which itself is 2.5\% of the brightness of HD 189733A.
Because the nearby stars are so faint, and because our analysis
technique isolates the spectrum of the target star in space and
references it in time by dividing the in-transit spectrum by the
out-of-transit spectrum, we expect any contributions of
nearby stars to the transmission spectrum of the planet HD 189733b
to be entirely negligible.

We needed to scan at $\sim$2\arcsec s$^{-1}$ to properly expose the detector
and to avoid saturation. Also we wanted samples often enough
to avoid overlapping the two spectra of the primary star HD 189733A
and the secondary star HD 189733B, 11.2\arcsec\ apart. Combined,
those requirements imply sampling every $\sim5$ s, or faster. Of MULTIACCUM sequences
currently available for WFC3, only the RAPID sequence sufficed.

The observational data are interleaved in time; 7 forward scan
differences, each a 0.85-s exposure, followed by a 51 second gap
as the scanning telescope reverses and the detector is prepared for
the next exposure, followed by the 7 reverse scan-differences each
one also 0.85-s long, and then a 51 second gap to return to the
beginning of the pattern and to prepare the detector for the next
forward-scan exposure. With two 6-s exposures every 114 s, the
observational duty cycle would be 10.5\%, although because only 9
of the 14 MULTIACCUM differences were useful, the duty cycle was
6.7\%. In total,
we obtained $16 \times 9\times 0.85 = 122$-s of exposure in transit,
which is only 3\% of the transit's duration. The latter 3\% is approximately half
of the 6.7\% observational duty cycle,
because the Earth occulted the target during the second half of the
transit. Rather unsatisfactorily, a 3\% duty cycle with the 2.4-m \hst\ would be equivalent
to observing at 100\% duty cycle
with a 0.4-m diameter telescope of the same end-to-end throughput.
Further improving WFC3's duty cycle by any practical operational changes would
be beneficial for observations of stars such as HD 189733b.
\section{Analysis}
\label{sec:analysis}

Two of us (N.C. and D.D.) analyzed the data nearly independently. 
As planned from the outset to maintain independence, no
computer code was shared between the two efforts, but both analysts had read or written
the relevant literature, particularly descriptions by \citet{MCC12b}
\citet{BER12}, \citet{DEM13}, \citet{KRE14a}, and \citet{KNU14}.
The two resulting exoplanetary
transmission spectra were compared and found to be very similar.
After discussions each analyst slightly revised his algorithms, which resulted in two
final spectra. In this section, we described one analysis,
that of N.C., the other method has been described already in its application to
HD 209458 and XO-1 \citep{DEM13}.
Table \ref{tbl-lc} lists the transit light curve of the N.C. analysis.
We describe the differences between the two
analyses at the end of this section and list the difference between the two resulting spectra
in the last column of Table \ref{tbl-spectra}.
The r.m.s. of the differences, 52 ppm, is 0.75 times the median Poisson noise associated with the 4-column binning,
indicating that the differences in the resulting binned spectra are statistically less significant than the physical limit
imposed by Poisson statistics.

We begin the analysis with the set of Intermediate MultiAccum Flexible Image Transport System (ima.fits) files,
which are produced by the CALWFC3 pipeline at STScI and have a number of corrections applied,
including dark current subtraction and nonlinearity correction \citep{RAJ10}.
As described in Section 2, we discard the data from the visit's first HST orbit.
We sort the files into two sets, forward scans (keyword POSTARG2 = 54.9) and reverse
scans (POSTARG2 = -2.0), and analyze each separately until combining them near the end of our analysis.
The separate analyses of forward- and reverse-scans are very similar, differing only in the parameters defining the useful
readouts and the pixels of interest, both of which differ slightly due to the forward and reverse scans not overlapping exactly.

From each 3-D multi-extension ima.fits file, we extract the difference images $\Delta I_i$ corresponding to the 
$i$th difference of consecutive nondestructive reads of the detector subarray.
We divide the 2-D difference images by a 2-D flat field obtained through the F139M filter. Because the
in-transit spectrum of HD 189733 will be divided by its out-of-transit spectrum to form the planetary transmission
spectrum, i.e. because the flat field is self-calibrated,
the choice of particular flat-field is unimportant. 
Flat-fielding makes bad pixels and cosmic ray hits easier to identify.
We perform such identification and accordingly correct the images by interpolation as described by \citet{CRO14}.

We define fixed regions on the 2-D images corresponding to the target's signal area and a corresponding sky area.
We subtract the sky's value from each image. Next we sum each column over the
stellar region, yielding a 1-D spectrum of HD 189733 for each difference image. 
We average those into
in-transit spectra (IN, from the fourth \hst\ orbit) and out-of-transit spectra (OUT, from three other \hst\ orbits).
We form each of nine planetary transit spectra, (OUT-IN)/OUT, associated with the nine difference images (four {\it forward} and five {\it reverse}),
which we then average to form a single planetary spectrum, subtract the wavelength-integrated mean, and bin every four columns
to produce the final, differential planetary spectrum (column 2 of Table \ref{tbl-spectra}).

We calibrated the wavelength as a function of
detector column by comparing the (unbinned) WFC3 spectrum of HD 189733 to the
spectrum of a K1 V comparison star \citep{RAY09}.\footnote{
http://irtfweb.ifa.hawaii.edu/$\scriptstyle\sim$spex/IRTF\_Spectral\_Library/Data/K1V\_HD10476.txt
}.
We adjusted the comparison star's spectrum to match HD 189733's
spectrum by multiplying the comparison spectrum by the telescope
and instrument throughput \citep{DRE14}, transforming it from energy units to
photo-electron units, smoothing it to 4.7 nm (1 column) resolution,
high-pass filtering it and the WFC3 spectrum to emphasize weak spectral features in each,
and finally, match the two spectra by linearly transforming the comparison star spectrum with an offset and dispersion that
together constitute our wavelength calibration, which we estimate to be
accurate to better than 4.7 nm (1 column).
Additional details of this procedure are in \citet{CRO14}. 

The position of the spectrum shifts slightly throughout each exposure in response to feedback from the fine guidance sensors.
Variations in the scan rate perpendicular to dispersion
introduce apparent photometric variations due to each row having a slightly shorter (longer)
exposure time because the telescope's scan rate is faster (slower) than nominal (cf. Figure 1 of \citealp{DEM13}).
Because such variations are achromatic, they do not affect the differential exoplanetary spectrum.
We also examined the shifts in telescope pointing parallel to dispersion by methods similar to those described in \citet{DEM13}. 
The peak-to-valley
variation of the shifts throughout the visit of program 12881 was typical for \hst\ \citep{GIL05}, $\sim$0.1 column ($\sim$0.014\arcsec), similar to that for XO-1 and much less
than the $\sim$1.0 column shift of HD 209458b, both observed in program 12181 (cf. Figure 4 of \citealp{DEM13}).
Adjusting the spectra of HD 189733 for its small ($\la 0.1$ column) shifts made negligible difference in the final
exoplanet spectrum compared to not making the adjustments.\footnote{The analysis of D. D. included such adjustments;
the analysis of N. C. did not.}

An important difference between the two analyses is their smoothing. The analysis of D. D. 
smoothed its spectrum with a Gaussian of FHWM=4 columns whereas the other one, of N. C., 
used a box-car smoothing of full width 4 columns. Purely as a consequence of the
choice of smoothing kernels, for an idealized, featureless, flat power spectrum of noise in the unbinned spectrum,
the ratio of the noise of the resulting smoothed
spectra should be ${\rm \sigma_{D.D.}/\sigma_{N.C.} = \sqrt{\pi/\sqrt{32ln2}} = 0.817}$.
Stated another way, in terms of noise power, the D. D. spectrum has an effective
bin width of 6 columns, or 1.5 times the N. C. spectrum's 4-column bin width.\footnote{To derive the factor, recall that the convolution of
a Gaussian is another Gaussian in the Fourier-transformed space, and via Parseval's theorem, noise power is equal in both one space and
the space of its Fourier conjugate. We also verified the value of 0.817 by numerical simulation.}
The uncertainties in Table \ref{tbl-spectra} match this expectation.
\section{Results}
\label{sec:results}

\subsection{Light Curve}

\begin{figure}
\plotone{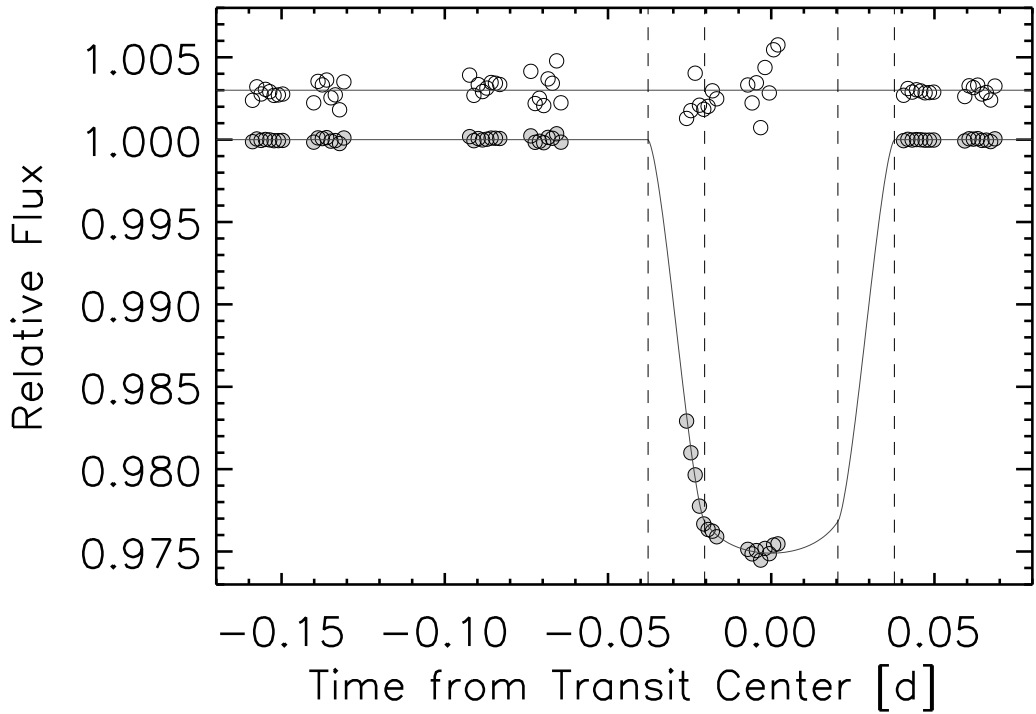}
\caption{The white-light transit curve of HD 189733b, integrated
over the WFC3 G141 bandpass with 0.8-seconds exposure per point.
We obtained a total of nine such light curves, nearly simultaneously,
via the MULTIACCUM samples of the detector.  A limb-darkened transit model
is overlaid (solid line) and the four planetary
contacts are indicated (vertical dashed lines). Residuals have
been expanded five-times in vertical scale and shifted above unity
for clarity.  Instrumental artifacts has been removed by fitting
out-of-transit data (see text). The first HST orbit of the single
visit is not shown.
\label{fig2}}
\end{figure}

We subdivided the available WFC3 data into nine
separate wavelength-integrated ``white light'' curves, corresponding to the nine differences between
MULTIACCUM samples in which HD 189733 was on the subarray. 
Figure \ref{fig2} illustrates one example of the nine available.
We removed trends in each light curve as follows,
for each of the nine MULTIACCUM differences separately.
We fit a straight line by least-squares to the first points of the
second, third, and fifth \hst\ orbits, and divided the first point
of each \hst\ orbit (including the in-transit fourth orbit) by the fit.
We repeated the procedure for the second point, and so on, for all
thirty-two scans per HST orbit. Table \ref{tbl-lc} lists the nine light
curves, both before and after de-trending. 

From the light curve of HD 189733b's
transit we obtain the mean level of the transit depth $R_p^2/R_s^2$,
which is important for comparison with measurements at other
wavelengths from other instruments. Otherwise, however, we are not
as interested in the light curve as the differential transit spectrum described in the next section.
We adopted HD 189733b's
ephemeris, orbital inclination, and normalized semi-major axis from
\citet{KNU12}. We fit the light curve with the procedure of
\citet{MAN02}, using quadratic limb darkening coefficients $u_1 =
0.1378$ and $u_2 = 0.232$, which we derived by fitting two parameters
of a quadratic law to the three-parameter limb-darkening law adopted
by \citet{SIN09} for $\lambda = 1.66$ \um.  The resulting mean
transit depth, appropriate for the integrated WFC3 bandpass, is
$R_p^2/R_s^2 = 0.024332\pm0.0001$, where the estimated uncertainty
(100 ppm) is the standard deviation of the fits to the nine light
curves. The observed value of $R_p^2/R_s^2$ is an upper bound in so
far as it neglects unocculted star spots (cf. \S\ref{sec:spots}).

\subsection{Exoplanet Transmission Spectrum}

\begin{figure}
\plotone{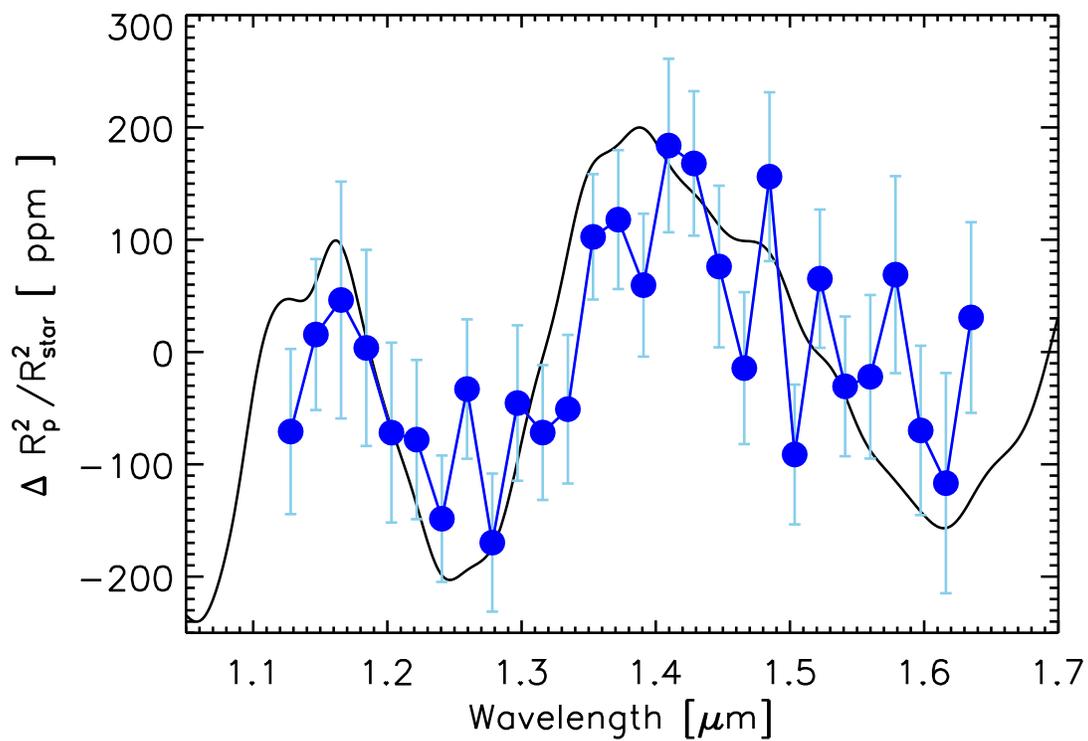}
\caption{Transmission spectrum of the exoplanet HD 189733b compared to a
theoretical model of a clear planetary atmosphere of solar composition. The model spectrum has a vast number of water vapor lines; for clarity
the model spectrum has been Gaussian smoothed to FWHM=0.03 \um.
\label{fig3}}
\end{figure}

The exoplanet transmission spectrum is the transit depth $R_p^2/R_s^2$ as a
function of wavelength. 
In column 2 of Table \ref{tbl-spectra} we report the differential transmission spectrum as (IN-OUT)/OUT,
with its mean value subtracted, where IN is the mean in-transit spectrum
(from the fourth \hst\ orbit) and OUT is the mean out-of-transit spectrum
(from the other \hst\ orbits, neglecting the first one).
Figure \ref{fig3} illustrates the spectrum extending
from 1.13 \um\ to 1.64 \um\ in 28 bins. Each bin corresponds to 19 nm or
four detector columns. The simple (IN-OUT)/OUT estimate neglects limb darkening. Our more complicated analysis \citep{DEM13}
involves subtracting a scaled template spectrum from each exposure's 1-D spectrum and then fitting
a transit light curve to the residuals to derive the differential transit depth at each wavelength.
In the latter method, limb darkening can be included or not, and we experimented with either choice.
Although the mean ``white light'' transit depth is sensitive to
the limb-darkening correction, applying a wavelength-dependent limb-darkening correction, or not,
made negligible difference to the differential transit depth. To emphasize the latter point,
we computed one differential spectrum with no limb-darkening correction at all (column 2 of Table \ref{tbl-spectra}),
although we did model limb-darkening for the white-light transit depth in the previous section.

We estimated the uncertainties in column 3 of Table \ref{tbl-spectra} 
as follows. We formed an individual 1-D stellar spectrum
from each difference image. We form a template spectrum from data
taken out-of-transit. We then scale and subtract the
template from each of the individual spectra in order to minimize the
r.m.s. of the resulting residuals. For each spectral channel, we group the residuals into two sets,
one in-transit and one out-of-transit, divide by the mean flux, and evaluate their respective
standard deviations $\sigma_{in}(\lambda)$ and $\sigma_{out}(\lambda)$. We combine the
latter in quadrature, accounting for the number of measurements in-transit or out-of-transit
respectively, in order to estimate the uncertainties of each spectral channel, $\sigma(\lambda)$.

\begin{figure}
\epsscale{1.20}
\plotone{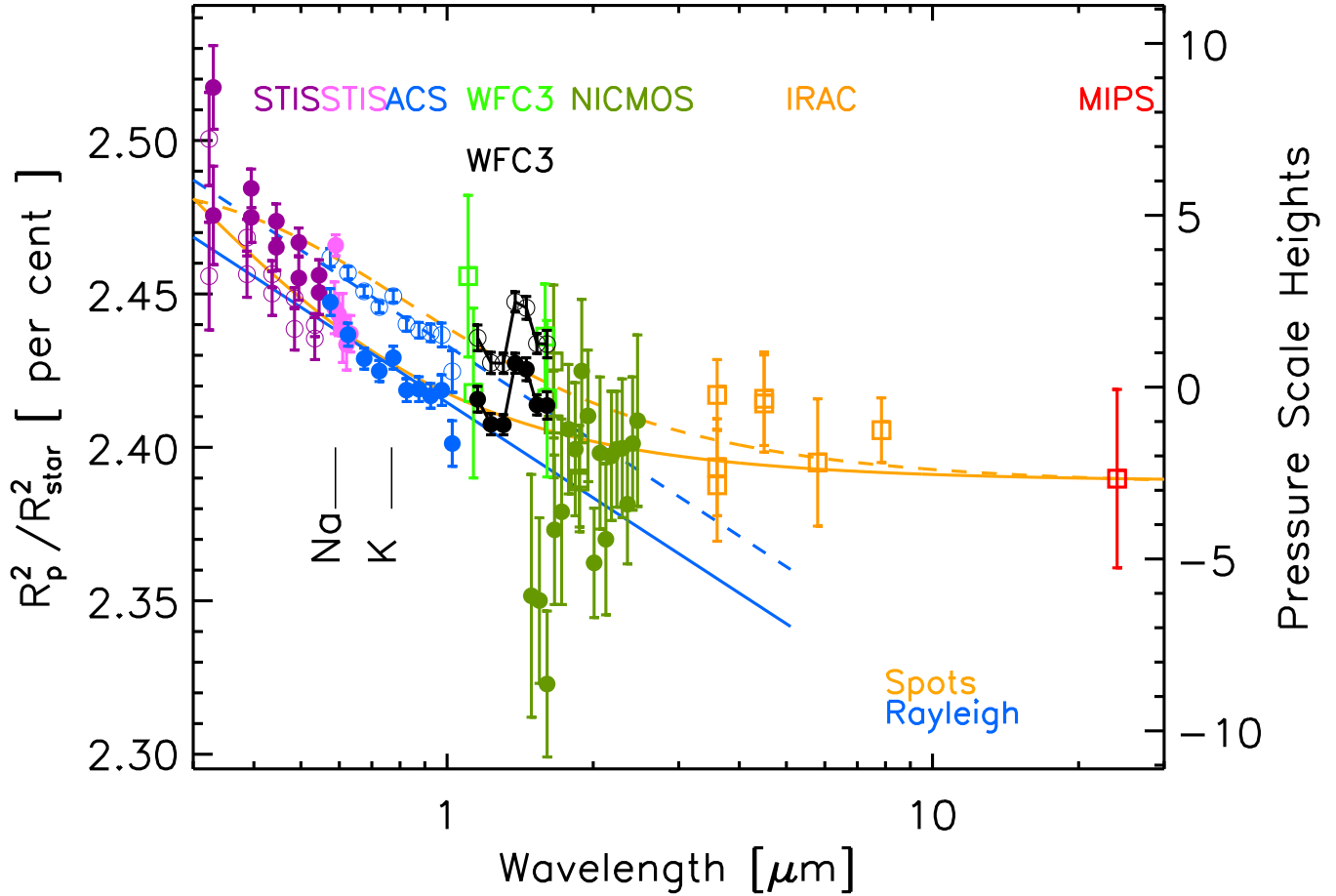}
\caption{Transmission spectrum of the exoplanet HD 189733b. 
{\bf Observations:} 
Data from Table \ref{tbl-spectra} of this work, binned to seven points, are indicated by the filled black circles connected by a line. 
The WFC3 spectrum's S-shaped undulation is attributed to water vapor (cf. Figure \ref{fig5}).
{\bff Except for the WFC3 spectrum, all of the data and associated uncertanties are from Table 5 of \citet{PON13}, including}
spectroscopy (filled colored circles) and photometry (open colored squares).  
{\bff As such, the ilustrated NICMOS spectrum was reported originally by \citet{GIB11} and has a median uncertainty per point of 234 ppm.
The median uncertainty per point estimated by three different analyses of those same NICMOS data
ranges from 80 ppm to 234 ppm, although the shapes of the independently-derived spectra are similar \citep{SWA14}.}
Data from Table 3 of \citet{SIN11} include five pairs of points from the G430L grating of STIS (open violet circles),
except their wavelengths have been reduced 2\% for clarity. 
Data from Table 2 of \citet{PON08} include ten points from the G800L grating of ACS (open blue circles).
Different corrections to the same data account for the different transit depths (open vs. filled circles; $\lambda < 1.7 \mu$m).
Uncertainties are represented by $1-\sigma$ error bars.
Transitions of atomic sodium and potassium are indicated.
{\bf Models:} 
The blue dashed line is a fit to the \citet{PON08} data (open blue circles) by \citet{LEC08a} with a model
of a clear, Rayleigh-scattering planetary atmosphere of the form of Eq. 1 of \citet{LEC08b}.
The blue solid line is the same fit, shifted down to match the data of \citet{PON13} (filled blue circles).
The orange lines have a fixed planetary radius $R_p$ and black-body stellar models with unocculted star spots.
The temperature of the stellar photosphere is 5000 K.
The modeled spot temperatures and covering fractions are 4600 K and 0.08 (solid), and 3700 K and 0.056 (dashed).
Both spot models have been normalized at $\lambda = 24\mu$m to match the MIPS data point.
An annulus of radius equal to $R_p$ and width equal to a single T=1200 K pressure scale height ($H=193$ km) corresponds to 112 ppm of
transit depth, as indicated by the right scale.
\label{fig4}}
\end{figure}

\begin{figure}
\epsscale{1.20}
\plotone{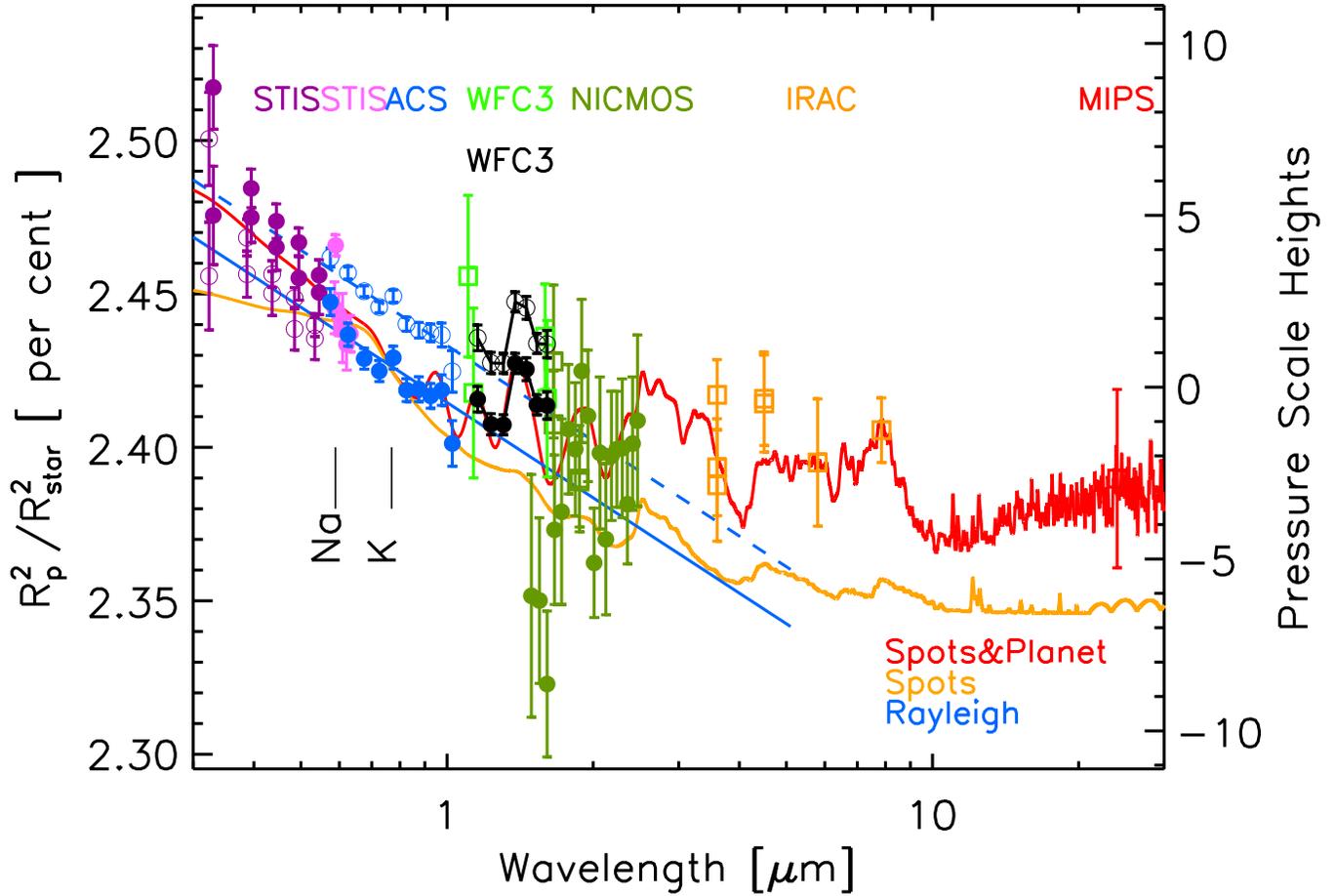}
\caption{Transmission spectrum of the exoplanet HD 189733b. 
Same as Figure \ref{fig4} except PHOENIX atmosphere models replace the black body models and 
the planetary radius $R_p$ varies according to a physical model (see text).
{\bf Observations:} 
The upper WFC3 spectrum from our analysis (black open circles) is as-observed; the mean transit depth for the WFC3 data
is approximately at the same level as the $\sim$1 \um\ end of the ACS data (blue open circles) reported by \citet{PON08},
which had been corrected for an assumed unocculted star spot level of 1\%.
The lower WFC3 spectrum (black filled circles) has been shifted down by 300 ppm to better
match the ACS data (blue filled circles) corrected for an unocculted star spot level of 1.7\% by \citet{PON13}. 
{\bf Models:} 
One model (upper, red line) combines two effects: 1) unocculted star spots with temperature $T(spot)=3700$ K and spot fractional area $\delta = 0.056$,
and 2) a clear planetary atmosphere of solar composition, a mixing ratio for water of $5\times 10^{-4}$, and zero alkali metal lines (Na and K) for a gas giant planet with physical parameters commensurate with HD 189733b.
The other model (lower, orange line) is solely the contribution of the unocculted star spots of the first model.
Both models have been smoothed with a Gaussian of FWHM=0.089 \um\ for clarity.
\label{fig5}}
\end{figure}

In Figure \ref{fig4} we bin the data from column 2 of Table \ref{tbl-spectra} into seven points, with a channel
width of 75 nm (16 detector columns). With that smoothing, the local maxima
of the transit depths in the 1.15 \um\ and 1.4 \um\ water vapor features occur at the first and fourth data points,
with a local minimum, the third point, in between. From those, we estimate the amplitudes of the 1.15 \um\ and 1.4 \um\ features
respectively are $83\pm 53$ ppm and $200\pm 47$ ppm greater than
the local minimum at 1.3 \um. The stated uncertainties are the quadrature sum of the uncertainties of the two points in each
case. The 1.4 \um\ feature in particular is detected at much greater statistical significance than $\sim4\sigma$, i.e.
200/47, because the feature extends over two 75-nm wide points and the adjacent baseline extends over two or four points
instead of just one. By averaging over four points, two inside the 1.4 \um\ feature and the two adjacent to it, the $\sim4\sigma$
result becomes $\sim8\sigma$, commensurate with the feature's appearance in Figures \ref{fig4} and \ref{fig5}.

{\bff The peak-to-valley amplitude of the 1.4 \um\ feature in Figure \ref{fig3}, which has spectral resolution of 30 nm FWHM in
the model or 19 nm (4 columns) in the binned data, is $\sim$400 ppm or $\sim$350 ppm, respectively. The peak-to-valley measurement
of a noisy spectrum depends on the smoothing that reduces noise and hence reduces the height of the peak and the depth of
the valley. 
The similarity of the model to the data in Figure \ref{fig3} illustrates that the data are consistent with a clear-atmosphere of solar-composition, with
a volume mixing ratio for water of $\sim5\times10^{-4}$ and a pressure scale height commensurate with a temperature of 700 K. The 
$\sim5\times10^{-4}$ mixing ratio has been used in prior models as well (e.g. \citealp{TIN07}, \citealp{SWA08}, \citealp{DAN14}) and is
consistent with the broad range of values presented in Figure 3 of \citet{SWA14}, $\sim1\times10^{-4}$ to $\sim1\times10^{-2}$.
However, inspection of Figure \ref{fig3} shows that slightly reducing our model's amplitude would improve its fit to the data. Hence, either
the volume mixing ratio or the temperature should be smaller than the assumed values. \citet{MAD14} explore the fitting of these data
in greater detail and conclude that the H$_2$O volume mixing ratio is sub-solar. 
}

\section{Star Spots}
\label{sec:spots}

HD189733 is an active star (Boisse et al. 2009), and transit
observations exhibit ubiquitous star spot crossings (Pont et al.
2007).  Our WFC3 observations have only partial coverage during transit
because of occultation by the Earth, and no spot crossings are obvious in
our data. In any case, because HD 189733 is a spotted star, we must address two effects {\it unocculted} star spots may have on the transmission spectrum.
First, unocculted star spots can make the radius of a transiting planet appear larger in the blue than in the red, because 
unocculted spots make the star darker and redder than it would {\bff appear} otherwise. If in fact a star is {\it darker} away from the transit chord than a model assumes,
that model will overestimate the size of the transiting planet. If in fact the star is also {\it redder} away from the transit chord than the model assumes, then the
overestimate of the planet's size will be even greater in the blue than in the red. 
Second, at $T_{eff} = 5000 $K, HD 189733's photosphere is too hot for water to exist, but
in very cool star spots, water vapor could exist. Water vapor absorption in unocculted star spots will make the star spots darker in spectral features
of water than in the continuum, and hence make the radius of the transiting planet appear larger in water-vapor features than it would in the continuum \citep{DEM13}.
Quantitatively and specifically for HD 189733, we show in this section that the first effect is likely to be significant but the second effect is not.

\citet{PON13} analyze star spots on HD 189733 and their potential effects on the transmission spectrum.
In their analysis, they show that star spots modulate the stellar flux primarily due to stellar
rotation, and also due to the changing set of spots and their characteristics. 
However, latitudinal {\it bands} of star spots would not modulate the flux much, because as one spot rotates into view,
another one rotates out of view.\footnote{From observations of the Rossiter-McLaughlin effect for HD 189733b, we know that the stellar rotation axis
is in the plane of the sky \citep{TRI09}.}
If HD 189733 has a large polar spot, or one or more latitudinal band(s) of spots beyond the latitudes at which the
transit chord passes, then those unocculted spots will contribute to the apparent rise in the ``planetary'' transmission spectrum seen in the visible
with ACS \citep{PON08} and in the ultraviolet with STIS \citep{SIN11}.
\citet{PON13} were aware of such possibilities but chose to interpret that rise in the spectrum as evidence of Rayleigh-scattering
dust in the planet's upper atmosphere instead of spots on the star. In this paper we match all of the transit data
assembled by \citet{PON13} and the WFC3 spectrum derived here with a
combination of a clear planetary atmosphere and a spotted stellar photosphere.

As in prior work (e.g., Pont et al. 2008, 2013; Sing et al. 2011), we model star spots
in aggregate and ignore limb darkening. 
In the spotted-star model, a fraction $\delta$
of the star's projected surface area is covered with spots, 
emitting with radiance $F_\nu(spot)$, less than that of the photosphere, $F_\nu(phot)$.
Such a model predicts an apparent transit depth,
\begin{equation}
{{\tilde{R}_p^2}\over{\tilde{R}_s^2}}    = {{R_p^2}\over{R_s^2}} {1\over {1-\delta (1-F_\nu(spot)/F_\nu(phot))}}, \label{eq-r-tilde}
\end{equation}
where the first factor on the right is the transit depth that would be measured in the absence of
star spots.

\citet{AIG12} used simulated, rotating spotted stars modeled on HD 189733
to estimate the product of its spots' area coverage $\delta$ and (1-contrast),
\begin{equation}
f = \delta {\rm (1-contrast)} \approx {{\psi_{max} - \psi_{min} + \sigma}\over{\psi_{max}+\sigma}}, \label{eq-f}
\end{equation}
where $\psi_{max} - \psi_{min}$ is the difference in the stellar flux from
maximum to minimum, and $\sigma$ is the scatter in the light curve, also due to rotational modulation (not observational uncertainty).
From six years of photometry at $\lambda = 0.51$ \um\ of HD 189733 illustrated in Figure 3 of \citet{PON13}, we estimate $\psi_{max} - \psi_{min} \approx 0.04 \psi_{max}$ and
hence $f$ can be comparable to, or larger than, $0.04$.
Because $f$ represents the product, spot area times (1-contrast), and contrast must be less than unity,
in general, the fraction $\delta$ of the star's disk covered with spots must be greater than $f$.
Thus, for HD 189733, we expect $\delta$ at times is greater than approximately $0.04$.
A caveat worth noting is that the analysis of \citet{AIG12} is most appropriate for stars for which
a single active region dominates the photometric variability. Also, as noted by \citet{AIG12}, geometric
{\bff foreshortening de-emphasizes the contribution to the rotational modulation of spots near the stellar limb.}
With those caveats in mind, we admit the
possibility that Equation \ref{eq-f} underestimates $f$ and hence $\delta$ could be even larger than 0.04,
if for instance HD 189733 has many active regions spread uniformly in longitude, i.e. bands of star spots.

{\bff The values of \citet{PON13}, which we use as a starting point in our analysis,
have been corrected already for unocculted star spots that
modulate the light curve, with time-dependent ``AC'' values for $\delta$ corresponding to
flux changes in the optical between $-0.93$\% and $+1.47$\% (Table 3 of \citealp{PON13}).
The most important corrections are to the ACS and STIS data, because the transmission spectrum
at wavelengths longer than
$\sim1$ \um\ are less affected by the unocculted spot correction and/or by the Rayleigh-scattering
dust model (Figures \ref{fig4} and \ref{fig5}). The unocculted spot corrections derived by
the procedure of \citet{PON13} produced a discontinuity between the STIS and ACS transmission
spectra in the region of overlap at $\sim$0.5 \um. To eliminate the discontinuity, they 
applied a smaller correction than their method had indicated for unocculted spots at the epoch
of the STIS observations.
While \citet{PON13} have corrected for an AC component, here we
investigate an underlying ``DC'' component and its potential effects on the transmission spectrum.
The DC component is an additional, hypothetical set of unocculted spots 
in bands or a polar cap.  With either such configuration, they would not modulate the light curve
or precision radial velocities either.
Polar spots have been deduced from Doppler imaging of rapidly rotating stars, e.g. the K0
dwarf AB Dor has a rotation period of 0.5 days and exhibits a prominent, long-lived polar spot
\citep{CAM95}, but we know little about the spots of more slowly-rotating stars such as HD 189733.
}

\subsection{Black Body models}
\label{sec:bb}

Before proceeding to the more realistic PHOENIX atmosphere models, we begin with
the approximation of black bodies for the stellar photosphere and
the star spots, at temperatures $T_{phot}$ and $T_{spot}$ respectively. In this approximation, Equation \ref{eq-r-tilde} becomes
\begin{equation}
{{\tilde{R}_p^2}\over{\tilde{R}_s^2}}  = {{R_p^2}\over{R_s^2}} {1\over {1-\delta (1-{{e^{h\nu/kT_{phot}}-1}\over{ e^{h\nu/kT_{spot}} - 1     }})}},
\end{equation}
where $h$ is Planck's constant, $k$ is Boltzmann's constant, and $\nu$ is the observing frequency.
The spots' multiplicative effect on the apparent transit depth as a function of wavelength asymptotically approaches
one value in the Wien limit (in the ultraviolet) and another in the Rayleigh-Jeans limit (in the far infrared):
\begin{eqnarray}
{{\tilde{R}_p^2}\over{\tilde{R}_s^2}} &=& {{R_p^2}\over{R_s^2}} {1\over {1-\delta}}, ~~~~~~~~~~~~~~~~~~~~~~~~~~~~~~Wien\\
                                      &=& {{R_p^2}\over{R_s^2}} {1\over {1-\delta (1-T_{spot}/T_{phot})}}, ~~~~Rayleigh-Jeans.
\end{eqnarray}
The difference between the two limits is 
\begin{eqnarray}
\Delta{{\tilde{R}_p^2}\over{\tilde{R}_s^2}}   &=& {{R_p^2}\over{R_s^2}} ( \delta {{T_{spot}}\over{ T_{phot} }} + O(\delta^2)),\\
                                              &\approx& 960 {{\delta}\over{0.04}} {{T_{spot}}\over{ T_{phot} }} {\rm ~~~ppm}, \label{eq-delta}
\end{eqnarray}
where the first expression is generic and the second is specific for HD 189733b's nominal transit depth of 2.40\%.
Intuitively, the greater the fractional area of star spots, $\delta$, the greater the increase in apparent
transit depth in the ultraviolet relative to the thermal infrared. Counter-intuitively (perhaps), the cooler the spots,
the less their effect on the difference in apparent transit depth $\Delta{{\tilde{R}_p^2}\over{\tilde{R}_s^2}}$.
This is because in the Wien limit, the radiance of cool spots is arbitrarily smaller than that of the photosphere,
i.e. the spots are essentially black silhouettes, whereas in the Rayleigh-Jeans
limit, radiance is proportional to temperature, so the {\it difference} in radiances (Rayleigh-Jeans minus Wien) is
proportional to the ratio of temperatures, and then so is the difference in apparent transit depth.

These characteristics of blackbodies inform our expectations for the effect of unocculted spots on the apparent transit depth.
Pont et al.  (2008) estimate that $T(spot) \sim$4000K, and \citet{SIN11} find $T(spot) = 4250\pm250$ K for HD 189733.
If the typical spot covering fraction $\delta$ is approximately 4\%, then Equation \ref{eq-delta} predicts the 
difference in apparent transit depth between the ultraviolet and the infrared should be 
$\sim800$ ppm, entirely due to unocculted star spots and not due to any variation in planetary radius
with wavelength. From Table 6 of \citet{PON13}, the observed difference in transit depth between 0.33 \um\ and 24 \um\ is 1100 ppm.
Evidently, a majority fraction of the increased apparent transit depth in the ultraviolet compared to the infrared
could be due to {\it more} unocculted star spots than that assumed by \citet{PON13}.
If so, then there would be no need for Rayleigh-scattering, non-absorbing dust
in the atmosphere of HD 189733b in order to match the slope in the apparent transit depth with wavelength from 
0.3 \um\ to 1.0 \um\ as in the interpretation of \citet{PON13}.

\subsection{PHOENIX models}
\label{sec:phoenix}

To achieve greater fidelity than the black body models permit,
we have modeled the effect of unocculted star spots using Phoenix
NextGen model atmospheres (Hauschildt et al. 1999).  We use a
5000/4.5/0.0 (Teff/log(g)/[M/H]) model for the star, and for the
spots we use NextGen models whose temperature we vary between 3000
and 5000K, in steps of 50 K via interpolation in the NextGen grid.
We constrain the log(g) and metallicity of the spots to be the same
as the stellar model, and we ignore center-to-limb effects on the
star spot spectrum.  This procedure is consistent with similar
previous work (e.g., Pont et al. 2008, 2013; Sing et al. 2011).
At each adopted spot temperature, we vary the assumed fractional
area covered by the spots, and we calculate the effect on the transit
depth as a function of wavelength (Equation \ref{eq-r-tilde}).

We calculate the $\chi^2$ of the
difference between the modeled transit depth as a function of
wavelength, using the observations in Table 5 of \citet{PON13}.
To start, we adopt the limiting-case in which all
of the observed planetary radius increase in the blue and UV is due to
unocculted star spots and the 
spot-free radius ratio $R_p/R_s$ is 0.15459, from the
longest wavelength transit measurement at 24\,$\mu$m (\citealp{KNU09}; \citealp{PON13}).  
In that case we find a best-fit spot
temperature of 4250K and a best-fit area coverage of 4.3\%.
This fitting process is well-posed in the
sense that the wavelength of the upward bend in the curve is determined by the temperature
adopted for the star spots (the cooler the spots, the longer the wavelength),
and the difference in apparent transit depth from the Rayleigh-Jeans end to the Wien end of the curves
is proportional to the product of the spots' temperature and area coverage (Equation \ref{eq-delta}).

Our best-fit temperature agrees well with spot temperatures inferred
from the amplitudes of spot crossings by this planet;
\citet{PON08} find that $T_{spot}\sim4000$ K, and \citet{SIN11} find $T_{spot} = 4250\pm 250$K.
Our best-fit model predicts $\delta = 0.043$, similar to
the value $f = 0.04$ that we estimated from the spot modulation of HD 189733's light curve at $\lambda = 0.51\mu$m.
In this case, the contamination of the 1.4 \um\ water-vapor feature is negligible, as we now discuss.

The model exhibits no spectral feature due to water
absorption in the unocculted star spots, since 4250K is too warm
for significant water absorption to be present.   Hence even the
worst-case assumption that all of the transit radius measurements
are due to star spots does not produce significant contamination
of the WFC3 water spectrum.  That remains true if a much cooler
spot temperature (3200K) is adopted, for example.
Such a spot is much cooler than the estimates noted in the previous paragraph,
and is also cooler than much larger spots seen
on other stars (O'Neal et al. 1998).  Moreover, even spots of T =
3200K do not produce sufficient water absorption to contaminate our
WFC3 measurement; the predicted amplitude of false water absorption is
$\sim$40 ppm - much less than
the $\sim$200 ppm feature at 1.4 \um\ that we measure.  We conclude that our WFC3 measurement
of water absorption is due to the planetary atmosphere, not to an
effect of unocculted star spots.

\section{Planet Atmosphere Models}
\label{sec:planet}

We constrain the H$_2$O abundance from our observed transmission
spectra of HD 189733b using model atmospheric spectra of its
terminator. We use the atmospheric modeling and retrieval technique
of Madhusudhan \& Seager (2009) and Madhusudhan (2012). The model
involves computing line-by-line radiative transfer in a 1-D
plane-parallel atmosphere, assuming hydrostatic equilibrium. The
temperature profile and chemical composition are free parameters
in the model, and, as such, has no constraints of radiative or
chemical equilibrium. This modeling approach allows one to compute
large ensembles of models ($\sim10^6$), and explore the parameter
space of molecular compositions and temperature structure in search
of the best-fitting models. Typically, the model includes opacity
contributions from the major molecular species expected in hot
Jupiter atmospheres (i.e. H$_2$-H$_2$ collision-induced absorption
and line absorption due to H$_2$O, CO, CH$_4$, and CO$_2$), though
H$_2$O is the most dominant molecule spectroscopically in the WFC3
bandpass. The transmission spectrum is less sensitive to the detailed
temperature structure, compared to a thermal emission spectrum, but
does depend on the characteristic temperature in the upper-atmosphere
above the infrared photosphere when viewed through the terminator.
In the present work, we assume a characteristic 1-D temperature
profile that has been derived from thermal emission measurements
in the past (Madhusudhan \& Seager 2009), and explore the range of
chemical compositions that explain the transit data.

We find that our WFC3 transmission spectrum of HD~189733b can be
explained very well by H$_2$O absorption at the terminator. The
WFC3 spectrum exhibits an absorption feature at 1.4 \um\ with an
amplitude of $\sim200$ ppm that coincides with the expected features
of gaseous H$_2$O in the same band. As shown in
Fig.~\ref{fig3}, the data can be fit well by a
solar composition atmosphere in chemical equilibrium. Such an
atmosphere has a H$_2$O mixing ratio of $\sim5\times10^{-4}$, where
the mixing ratio is defined as number density relative to H$_2$.
At the pressures probed by the observations ($P \lesssim 0.1$ bar),
the temperature profile has typical isothermal temperature of
$\sim$700 K, slightly cooler than prior P-T curves used to model NICMOS
transmission spectra of HD 189733b's terminator and its day-side emission
spectrum from {\it Spitzer} (c.f. Figures 5 and 8 of \citealp{MAD09}).
In principle, allowing for higher temperatures (e.g.
1000 K) leads to larger scale heights thereby constraining the
H$_2$O abundance to be slightly sub-solar {\bff ($\lesssim 10^{-4}$, \citealp{MAD14}).}
Nevertheless, the evidence of an absorption feature in the H$_2$O
band is unambiguous, despite the degeneracy between the
characteristic temperature and the H$_2$O mixing ratio. A featureless
flat spectrum does not fit the data, ruling out the possibility of
thick clouds fully obscuring the atmosphere observable along the planet's terminator.
On the other hand, a fully-obscuring cloud deck at lower altitudes (i.e. higher pressures,
$\gtrsim 0.1$ bar) in the atmosphere is not ruled
out by these WFC3 data.

\section{Combined Planetary Atmosphere and Star-spot Models}
\label{sec:combined}

Combining our WFC3 spectrum of HD~189733b with previously reported
spectra in other bandpasses reveal new constraints on its atmospheric
composition. The sum-total of previous data have led to an interpretation
of hazes or clouds in the atmosphere of
the planet (Pont et al. 2008; Sing et al. 2011; Pont et al. 2013).
In Figure \ref{fig5} we present a model that combines the effects
a clear planetary atmosphere and unocculted star spots.
For the
gas giant planet, we model a clear atmosphere of solar composition,
a mixing ratio for water of $5\times 10^{-4}$, and zero alkali metal
lines (Na and K). 
The Rayleigh scattering of H$_2$ in the
planetary atmosphere contributes significantly at wavelengths less than
0.5 microns. Incidentally, that scattering may be responsible for the higher geometric albedo
for the day side of HD 189733b observed at short wavelengths, $\lambda\lambda < 0.45$\um,
than at longer wavelengths, $\lambda\lambda > 0.45$\um\ \citep{EVA13}.
Because the STIS spectrum ($\lambda\lambda < 0.5$\um) has
a Rayleigh-scattering slope, the star spot contribution must saturate in order
that the model spectrum not rise too quickly into the ultraviolet.
That is accomplished with a cool, 3700 K spot temperature.
The star's model is a PHOENIX atmosphere model for the
5000 K stellar photosphere with unocculted spots of temperature
$T(spot)=3700$ K and spot fractional area $\delta = 0.056$. 
Although it fits the available data well, the
combined model presented here is not unique: it is one example of
many possibilities.  
As another example that could fit the transmission spectrum of HD 189733b is an ``enhanced Rayleigh'' model
proposed for WASP-12b by \citet{SIN13} in which the planet's atmosphere has both Rayleigh
scattering from aerosols and water vapor absorption.
A main purpose of this paper is to emphasize that fitting any transmission spectrum
with a planetary atmosphere model must also consider the potential
contribution and complication of unocculted star spots.

In prior work, the case for haze/clouds in the atmosphere of the planet has been
based on three key arguments: (1) The monotonic rise in
the spectrum from the near-infrared to the UV consistent with strong
Rayleigh and Mie scattering, (2) the lack of strong Na or K features
in the optical spectrum, i.e. presumably masked by a thick haze
layer, and (3) the lack of striking $H_2$O absorption
in previous datasets. We are able to
explain the monotonic blue-ward rise in the transmission spectrum
by considering the effect of star spots on the measured transit
depths in the relevant wavelengths (Section \ref{sec:spots}).
Secondly, the lack of Na and K features in optical transmission spectra could
be explained by either a low metallicity in the atmosphere \citep{HUI12} 
and/or by condensation of Na and K either in the low-temperature upper
atmosphere or on the night side. The latter model is a distinct possibility,
given that HD~189733b is one of the least irradiated hot
Jupiters ($T_{eq} \sim1200$ K). \citet{FOR05b} and \citet{MOR13} show the
condensation curves of various species in pressure-temperature phase space.
At $P\sim1$ bar, condensation temperatures of KCl and Na$_2$S are
$\sim800$K and $\sim1000$K, respectively, which can be higher than
the temperatures in HD~189733b at the corresponding pressure on the terminator.
Consequently, Na and K could be condensed out of the observable
atmosphere by arbitrarily large factors into their corresponding compounds (e.g. KCl, Na$_2$S).
Finally, the $\sim200$ ppm amplitude of the 1.4 \um\ feature reported here (and also
in other exoplanets; see \S1)
explains why it was not as obvious in previous datasets as had been anticipated.

\section{Summary}
\label{sec:summary}

We report detection of two water vapor features in the transmission
spectrum of HD 189733b, a strong one at 1.4 \um\ and a weaker one
at 1.15 \um.  Their shapes and amplitudes are matched well by a
solar-composition planetary atmosphere with a water mixing ratio
of $5\times10^{-5}$.
We investigate the possibility that the water vapor could exist in
star spots, but even for very cool spots ($T = 3200$ K), the amplitude
of the predicted 1.4 \um\ feature is much too small to produce the
observed feature.

The ensemble, polychromatic transit spectrum of HD 189733b, from
the ultraviolet to the thermal infrared, has been interpreted by
\citet{PON13}, and references therein, as evidence for Raleigh
scattering in the planetary atmosphere.  That interpretation requires
specific adjustments for unocculted star spots in order to match
the transmission spectrum across the entirety of the ultraviolet
and visible.  We re-interpret the polychromatic transit spectrum
using a clear planetary atmosphere and unocculted star spots.  In
the model presented here, Rayleigh scattering in a clear planetary
atmosphere contributes significantly to the slope in the planetary
radius as function of wavelength for 0.3\um$< \lambda < 0.5$\um.
In the range 0.5\um\ $< \lambda < 1.0$\um, the observed slope is
caused by unocculted star spots, not the planetary atmosphere.
Since Rayleigh scattering by transparent dust grains in the planetary
atmosphere has been central to prior interpretations of HD 189733b's
transmission spectra, our modeling of a clear planetary atmosphere 
and a spotted stellar atmosphere
is a significant revision in the interpretation. Given
the complexities of stitching together data from different instruments
observing transits of a variable star at different epochs, and given
the potential of omnipresent star spots to further confuse matters,
either the star spot interpretation or the Rayleigh scattering
interpretation, or some combination of the two, remain viable.
Prior work has rejected clear planetary atmosphere models based
upon the predicted large-and-broad alkali metal lines in transmission
spectra, Na I (0.589 \um) and K I (0.769 \um) \citep{FOR10}.  In
our model, the strengths of those features are reduced by arbitrarily
large amounts because Na and K are expected to precipitate out
(\citealp{FOR05b}; \citealp{MOR13}) at the low temperatures ($\sim$700
K) that we used to model the $\sim$200 ppm amplitude of the water vapor
feature at 1.4 \um.

\acknowledgments

The authors gratefully acknowledge everyone who has contributed
to the {\it Hubble Space Telescope} and the WFC3, and particularly those responsible for implementing the spatial scanning,
which was critical to these observations. We thank in particular John MacKenty and
Merle Reinhart. 
We acknowledge conversations with
{\bff
Suzanne Aigrain,
David Ciardi,
Suzanne Hawley,
Leslie Hebb,
Veselin Kostov,
Rachel Osten,
Frederic Pont,
Neill Reid,
and
David Sing.
}

This research used
NASA's Astrophysics Data System Bibliographic Services, and
the SIMBAD database, operated at CDS, Strasbourg, France, and
was funded in part by \hst\ grant GO-12881 and Origins of Solar Systems grant NNX10AG30G.

\begin{table}
\begin{center}
\caption{Summary of HST WFC3 observations \label{tbl-1}}
\begin{tabular}{lc}
\tableline
\tableline
 HST Program (P.I.) & 12881 (McCullough) \\
 Number of HST orbits & 5 \\
 Number of scans per orbit  &  16 Forward \& 16 Reverse \\
 Duration of scan (s) & 5.97 \\
 Scan rate (arcsec s$^{-1}$)[pixels s$^{-1}$] & (2.00)[16.5] \\
 Peak signal on detector (electrons per pixel) & $4.0\times 10^4$ \\
 Grism ($\lambda\lambda$) & G141 (1.1$\mu$m to 1.7$\mu$m) \\
 Detectory subarray size (pixels) & 512x512 \\
 Sample sequence & RAPID \\
 Samples per scan & 8 \\
 Start of first scan (HJD)  & 2456448.983024 \\
 Corresponding planetary orbital phase & -0.09948 \\
 Start of last scan (HJD)  & 2456449.272364 \\
 Corresponding planetary orbital phase &  0.03094 \\
\tableline
\end{tabular}
\end{center}
Notes: 
Forward and reverse scans were interleaved.
Planetary orbital phase is defined to be zero at mid-transit.
\end{table}

\begin{table}
\begin{center}
\caption{Light Curve\label{tbl-lc}}
\begin{tabular}{cccrcccc}
\tableline
\tableline
 EXPSTART MJD   & EXPSTART HJD  & Orbit & Scan & Sample & Flag & Counts & Flux \\
 2456448.541735 & 2456449.044294 &   2 &   1 &   1 &   0 &  55426193 &   0.99996 \\
 2456448.541745 & 2456449.044304 &   2 &   1 &   2 &   0 &  55452872 &   1.00014 \\
 2456448.541755 & 2456449.044314 &   2 &   1 &   3 &   0 &  55434284 &   1.00011 \\
 2456448.541765 & 2456449.044324 &   2 &   1 &   4 &   0 &  55335761 &   0.99988 \\
 2456448.542414 & 2456449.044973 &   2 &  -1 &   3 &   0 &  61415662 &   0.99990 \\
 2456448.542424 & 2456449.044983 &   2 &  -1 &   4 &   0 &  61552322 &   1.00005 \\
 2456448.542434 & 2456449.044993 &   2 &  -1 &   5 &   0 &  61495951 &   0.99995 \\
 2456448.542444 & 2456449.045003 &   2 &  -1 &   6 &   0 &  61571866 &   1.00017 \\
 2456448.542454 & 2456449.045013 &   2 &  -1 &   7 &   0 &  61610879 &   1.00067 \\
 2456448.543054 & 2456449.045613 &   2 &   1 &   1 &   0 &  55466554 &   0.99990 \\
 2456448.543064 & 2456449.045623 &   2 &   1 &   2 &   0 &  55489200 &   0.99993 \\
\tableline
\end{tabular}
\end{center}
Notes: 
The printed table is a truncated version of the electronic table, to illustrate the format.
Columns, left to right, are modified Julian date of the start of the exposure,
the associated heliocentric Julian date, the HST orbit in the visit, the scan direction
(1 = forward; -1 = reverse), the MULTIACCUM sample, a data-analysis flag, the total
number of photoelectrons from HD 189733, and the associated normalized flux after
detrending. Orbit 1 was not detrended and was ignored.
\end{table}
\clearpage

\begin{table}
\begin{center}
\caption{Transmission spectrum\label{tbl-spectra}}
\begin{tabular}{rrrrrrr}
\tableline
\tableline
 $\lambda$  & $\Delta R_p^2/R_s^2$ & $\sigma$ & Column &  $\Delta R_p^2/R_s^2$ & $\sigma$ & $\Delta\Delta$ \\
 ($\mu$m)  & (ppm) & (ppm) & & (ppm) & (ppm) & (ppm) \\
\tableline
 1.1279 &    -70 &     73 &     171.5 &    -58 &     55 &         -12 \\ 
 1.1467 &     15 &     67 &     175.5 &    -25 &     54 &          41 \\ 
 1.1655 &     46 &    105 &     179.5 &     36 &     53 &           9 \\ 
 1.1843 &      3 &     87 &     183.5 &     13 &     52 &          -9 \\ 
 1.2031 &    -71 &     80 &     187.5 &    -36 &     51 &         -35 \\ 
 1.2218 &    -77 &     70 &     191.5 &    -63 &     50 &         -14 \\ 
 1.2406 &   -148 &     56 &     195.5 &   -136 &     50 &         -11 \\ 
 1.2594 &    -32 &     62 &     199.5 &    -91 &     50 &          59 \\ 
 1.2782 &   -169 &     61 &     203.5 &   -108 &     50 &         -61 \\ 
 1.2970 &    -45 &     69 &     207.5 &    -21 &     49 &         -23 \\ 
 1.3157 &    -71 &     60 &     211.5 &    -86 &     49 &          15 \\ 
 1.3345 &    -50 &     66 &     215.5 &     20 &     49 &         -71 \\ 
 1.3533 &    102 &     55 &     219.5 &    106 &     49 &          -4 \\ 
 1.3721 &    117 &     61 &     223.5 &     75 &     49 &          42 \\ 
 1.3909 &     59 &     63 &     227.5 &     53 &     49 &           5 \\ 
 1.4096 &    183 &     77 &     231.5 &    195 &     49 &         -11 \\ 
 1.4284 &    167 &     64 &     235.5 &    165 &     50 &           2 \\ 
 1.4472 &     76 &     71 &     239.5 &     73 &     50 &           2 \\ 
 1.4660 &    -14 &     67 &     243.5 &    -41 &     50 &          26 \\ 
 1.4847 &    156 &     75 &     247.5 &     85 &     51 &          70 \\ 
 1.5035 &    -91 &     62 &     251.5 &     32 &     65 &        -123 \\ 
 1.5223 &     65 &     61 &     255.5 &     60 &     51 &           4 \\ 
 1.5411 &    -30 &     62 &     259.5 &    -53 &     52 &          22 \\ 
 1.5599 &    -22 &     72 &     263.5 &    -27 &     52 &           5 \\ 
 1.5786 &     68 &     87 &     267.5 &     61 &     53 &           7 \\ 
 1.5974 &    -69 &     75 &     271.5 &     12 &     54 &         -82 \\ 
 1.6162 &   -116 &     98 &     275.5 &   -111 &     55 &          -5 \\ 
 1.6350 &     30 &     84 &     279.5 &   -131 &     56 &         162 \\ 
\tableline
\end{tabular}
\end{center}
Notes. Units are as indicated; parts per million is abbreviated ppm.  
 The tabulated uncertainties apply to the differential transit depths; an
 additional uncertainty applies to the overall depth - see text. 
The first three columns refer to the analysis of N. C.; Columns 5 and 6 refer to the analysis of D. D.;
The last column contains the difference of the differential spectra, column 2 minus column 5.
\end{table}
\clearpage

\end{document}